\documentclass[amsmath, amssymb,nofootinbib]{revtex4}
\usepackage{graphicx}
\newcommand{\be}{\begin{eqnarray}}
\newcommand{\ee}{\end{eqnarray}}

\begin{document}
\author{R. Millo and P. Faccioli}
\title{CP-Violation in Low-Energy Photon-Photon Interactions}
\affiliation{Dipartimento di Fisica e I.N.F.N., Universit\`a degli Studi di  Trento\\
Via Sommarive 14 Povo (Trento), I-38100 Italy.}

\begin{abstract}
 CP-violation in the loop-mediated photon-photon interactions  would modify the birefringence of the quantum vacuum.
 We discuss the implications of this effect on an experiment in which light propagates in a Fabry Perot Cavity permeated by a time-dependent electric and magnetic field: for 
 some suitable orientation of the electric field, the effect does not cancel out in a round trip. We show that the violation of CP would imply the existence of characteristic Fourier components, in the intensity spectrum of the outgoing wave.  In order to estimate the magnitude of this effect within the Standard Model, we use chiral perturbation theory to compute the $\theta$-term contribution to the coefficient of the leading CP-odd vertex, in the low-energy effective field theory for photon dynamics.
\end{abstract}

\maketitle

\section{Introduction}

To date, CP-violations  have only been  observed in reactions in which the total flavor quantum number $F$ is not conserved ~ \cite{CPindirect, CPdirect, Babar, Belle}.
Such processes are understood in the Standard Model, in terms of the CKM flavor-mixing mechanism.

On the other hand,   CP-violation in flavor-conserving reactions has never been observed. Much of the experimental investigations have focused 
in the neutron electric dipole moment (NEDM), for which the most precise measurements give \cite{NEDMmeasure}
\be
|{\bf d}_n| < 2.9 \times 10^{-26} \textrm{ e}\cdot\textrm{ cm}. 
\ee

In the $\Delta F=0$ channel, the CP-violation generated by weak interactions is expected to be much smaller than in the $\Delta F\ne 0$ channels, because the charged weak currents do not contribute, at tree level.
For example, the weak contribution to the neutron NEDM is  only $d_{\textrm{weak}}\simeq 10^{-32} e\cdot  cm$ (see \cite{NEDMweak} and references therein).
On the other hand, all flavor-conserving  CP-odd matrix elements receive in principle a direct contribution from the so-called 
$\theta$ term of QCD, 
\be
    S_\theta = \theta~\frac 1{32 \pi^2} \int d^4x ~ F_{\mu \nu} \widetilde{F}_{\mu \nu},
    \label{Q}
  \ee
where $\widetilde{F}_{\mu \nu} = (1 / 2) ~ \varepsilon_{\mu \nu \alpha \beta} F^{\alpha \beta}$ 
is the dual gluon field strength  and $\theta$ is a (real) 
dimensionless parameter, which incorporates also a contribution form the weak sector,
\be
\theta= \theta_0 + \textrm{argdet}[\hat M],
\ee
where $\hat M$ is the complex, non-hermitian quark mass matrix.
Estimates combining the theoretical predictions of chiral effective Lagrangians with the existing  experimental
 bounds on the neutron NEDM have shown that the $\theta$ angle is in fact extremely small \cite{NEDMstrong},  
 \be
 \theta < 0.2 \times 10^{-10}.
 \ee  

The existing experimental results on the NEDM only provide information  on the strength of  CP-violating interactions inside hadrons.
On the other hand, they do not set constraints on the CP-violation in reactions in which the on-shell states do not contain neither quarks nor leptons.  In such a perspective, it has been studied the propagation of an electromagnetic wave, in an external magnetic field ${\bf B}$ \cite{Liao1,Rizzo}. We note that such a process may be in principle sensitive to exotic CP-odd microscopic dynamics, which is not constrained by the measurement of the NEDM~\cite{Liao2}. 
In a context of search for such a new physics, it is important to estimate the amount of CP violation in the same process, which is to be attributed to the Standard Model. The main goal of this paper is to compute the leading QCD contribution to such processes, which is induced by the QCD 
$\theta$-term.

\begin{figure}[t!]
\begin{center}
\includegraphics[width=8cm]{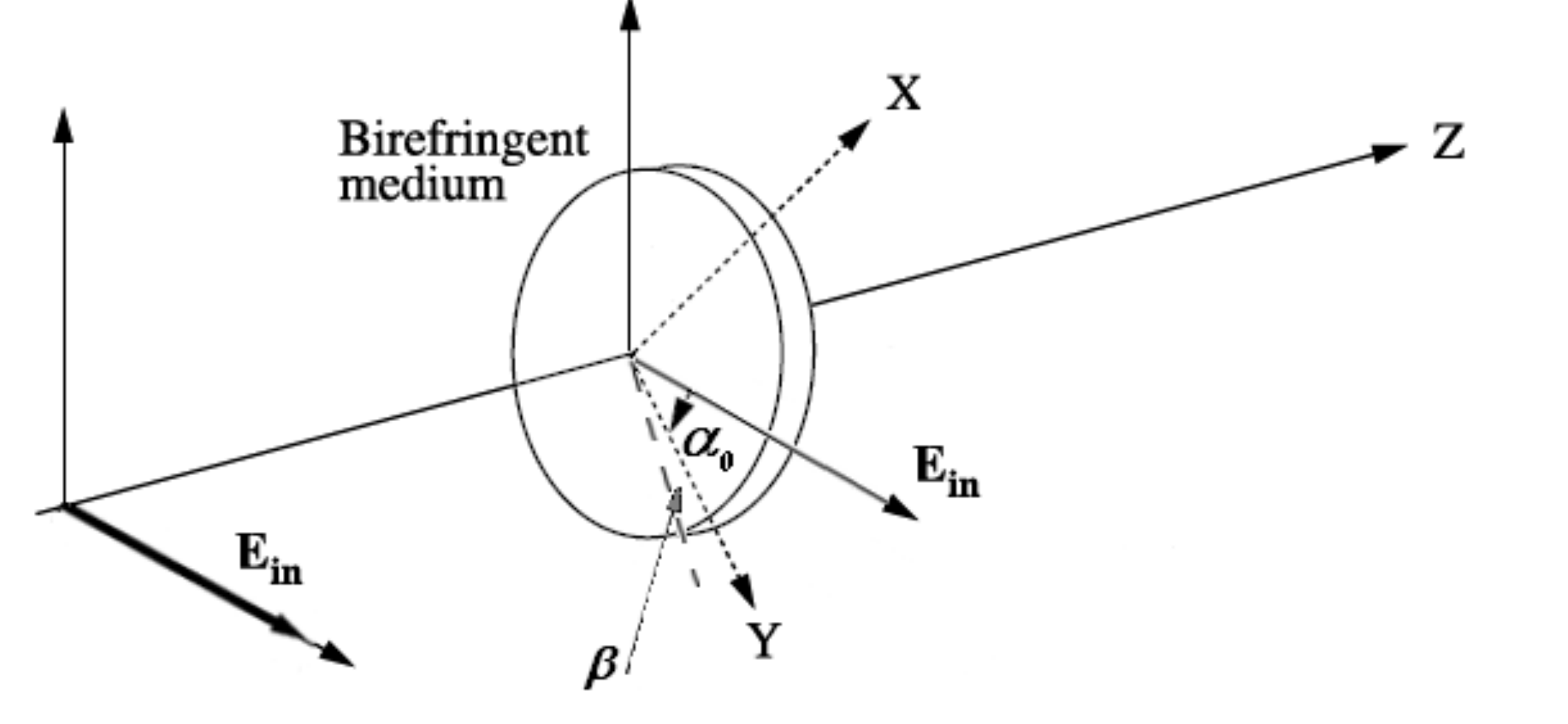}
\caption{Polarization planes and refraction indexes in an electro-magnetic wave propagating in a resonant cavity (see the details in the text).}
\label{default}
\end{center}
\end{figure}

In general, quantum fluctuations are known to be responsible for the non-linear response of the vacuum  to an external electromagnetic field. One of the  consequences of this effect is that a linearly polarized wave propagating in a region permeated by a magnetic field would acquire a finite ellipticity $\Psi$~(see e.g. \cite{Liao2,CPevenPsi} and references therein).  
In particular, if  CP symmetry is  conserved inside the loops mediating the quantum interactions of the electromagnetic wave with the external magnetic field, then the acquired ellipticity has the following general structure,
\be
\Psi_{\textrm{CP-even}}({\bf B}, L)= \frac{\pi}{\lambda}   (n_2({\bf B}) - n_1({\bf B})) ~L \times \sin 2 \alpha_0,
\ee
where $\lambda$ is the wavelength of the electromagnetic wave, $L$ is the distance travelled in the cavity,   $n_1$ and $n_2$ are the refraction indexes along the axes perpendicular and parallel to the external field and, $\alpha_0$ is the angle between the polarization axis of the incoming wave and the direction selected by the external field (see Fig. \ref{default}).

A small amount of CP violation inside the quantum loops mediating the light-by-light scattering would imply an additional contribution to the acquired ellipticity, in the form
\be
\Psi= \Psi_{\textrm{CP-even}}+ \Psi_{\textrm{CP-odd}}.
 \ee
 The correction from CP-odd term reads
\be\label{psiODD}
\Psi_{\textrm {CP-odd}} =  \epsilon~  \frac{\pi}{\lambda}   (n_2({\bf B})- n_1({\bf B})) ~ L \times \cos 2 \alpha_0, 
\ee
where $\epsilon$ is a small  parameter, which measures the amount of CP violation.

In order to  calculate explicitly the contribution to $\Psi_\textrm{CP-odd}$, coming from the  $\theta$-term of QCD we shall proceed as follows. 
First, we shall use the result of a microscopic calculation in chiral perturbation theory to  determine the effective coefficient of the leading  CP-odd operator, in the effective field theory for low-energy photon dynamics. Next, we shall express $\Psi_\textrm{CP-odd}$ in terms of such a coefficient. 
Our final result is
\be
\Psi_{\textrm {CP-odd}} \simeq  -2 \pi \frac{5}{9}\frac{\alpha^2 \theta}{m^2_{\eta'} (4\pi f_\pi)^2} \frac{L}{\lambda} {\bf B}^2 + \ldots,
\label{CPoddSM}
\ee
where  the ellipses denote higher orders, in the chiral and $1/N_c$ counting. 

The maximum contribution to the CP-odd ellipticity which can be attributed to the Standard Model is subject to the strict constraint on $\theta$ coming from the measurement of the NEDM. For a generic magnetic field 
orientation, it is extremely much smaller than the leading CP-even contribution to $\Psi$, which comes from QED.   
On the other hand, it is important to stress that the CP-even and CP-odd components of the ellipticity are qualitatively different and therefore can be experimentally separated. In fact, the CP-even part  vanishes for polarization angles $2 \alpha_0 =0, \pi$, i.e. in the configurations in which the CP-odd part is maximum.  Observing a significant excess of CP-odd ellipticity with respect to the estimate arising from Eq.~(\ref{CPoddSM}) would represent
 a signature of CP-violation coming from microscopic dynamics beyond the Standard Model, for example of the type investigated in \cite{Liao2}.  

In the last part of this work, we shall analyze in detail the possibilities of measuring the CP-odd ellipticity $\Psi_\textrm{CP-odd}$, in experiments in which the electromagnetic wave propagates in a Fabry-Perot cavity, such as PVLAS-Phase II \cite{PVLASII} or BMV \cite{Rizzo2}. These experiments make use of the additivity property of ellipticity, in order to increase the CP-even signal. On the contrary, one can show that after a reflection on a mirror, the coefficient $\epsilon$ in Eq.(\ref{psiODD}) changes sign. Hence the CP-Odd ellipticity $\Psi_\textrm{CP-odd}$ is not enhanced by propagating several times in the resonant cavity. Nonetheless, we shall show that if an electric field ${\bf E}$ is placed in the cavity together with the magnetic field ${\bf B}$, the total amount of CP-odd ellipticity after a single round trip is given by
\be
\Psi_{\textrm {CP-odd,R}}=\epsilon^{'}\,\frac{\pi}{\lambda}\,|{\bf B}|\,|{\bf E}|\,L\times\sin(\alpha_B+\alpha_E)\label{refl}
\ee
where $\alpha_B$ and $\alpha_E$ are the angles between the polarization axis of the incoming wave and the directions selected by the external magnetic and electric fields, respectively.\\
We shall also show that if the external fields are modulated, a finite amount of CP-violation would generate additional Fourier components in the spectrum of the intensity of the outgoing wave.

The paper is organized as follows. In section   \ref{EL} we introduce the effective Lagrangian for low-energy  photon-photon dynamics and we compute the QCD contribution to the 
effective coefficient of the leading CP-odd operator. In section \ref{compare} we study the birefringence of the vacuum  in the presence and in the absence of CP-violation, we review the calculation of the CP-even contribution to the ellipticity in QED and, we compute the CP-odd correction in QCD, using chiral perturbation theory and $1/N_c$ expansion.
In section \ref{nuovo} and \ref{pvlassect} we present our discussion of the possibility of CP-violation in experiments in which the electromagnetic wave is reflected by a mirror.
Conclusions are reported in section~\ref{concl}.

\section{Effective Lagrangian for photon-photon interaction with CP violation}
\label{EL}

At energy scales much below the electron mass, the quantum dynamics of the electromagnetic field can
 be rigorously formulated in terms of an effective theory, in which the only dynamical degrees of freedom are soft photons.  
One considers the most general effective Lagrangian compatible with gauge and Lorentz symmetries, which contains an infinite tower of  CP-even and CP-odd operators of increasing dimension,
\be
\mathcal{L}_{eff}=\sum_{i=1}^{\infty}~c_i(\Lambda)~\mathcal{O}_i,
\label{effL}
\ee
where $\Lambda\ll m_e$ is the cut-off scale. Perturbative power-counting rules imply that,  for momenta much below the cut-off, arbitrarily precise predictions can be obtained retaining only a finite number of lowest dimensional 
operators in (\ref{effL}).
The effective coefficients $c_i$'s in (\ref{effL}) encode the information about the quantum ultraviolet physics above the cut-off scale $\Lambda$. They can in principle be determined 
explicitly, by means of microscopic calculations in the underlying, more fundamental theory. 

Starting from the electromagnetic field stress tensor, one can construct only two Lorentz invariant terms: a scalar,
\be
X={\bf B}^2-{\bf E}^2=\frac{1}{2}F_{\mu\nu}F^{\mu\nu},
\ee
and a pseudo-scalar,
\be
Y={\bf B}\cdot{\bf E}=\frac{1}{2}F_{\mu\nu}\tilde{F}^{\mu\nu}.
\ee
Hence, the most general effective Lagrangian is constructed by combining  powers of such operators. Clearly, the CP-odd dynamics is encoded in the terms containing an odd number of $Y$ operators.
The lowest-order terms in the effective Lagrangian  read
\begin{equation}\label{EH}
\mathcal{L}_{eff}=-\frac{1}{2}X+ a~Y + b~X^2+c~Y^2 + d~X Y+ \ldots
\end{equation}

The effective coefficient $a$ can be set to zero, since the pseudo-scalar operator $Y$ is a total derivative, while
the effective coefficients $b,c$ and $d$ appear only at the quantum level. In principle, these coefficients receive contributions from the electro-weak and strong sectors of the Standard Model and, possibly, also from dynamics beyond the Standard
Model.  In particular, the leading contribution to $b$ and $c$ comes from QED and has long been calculated\cite{Heisemberg}
\be
b=\frac{2\alpha^2}{45m^4_e},\qquad c=7b.
\ee

The coefficient $d$ in Eq.~(\ref{EH}) parametrizes the short-distance CP-violating dynamics.
Within the Standard Model, if the QCD
$\theta$-angle is greater than $10^{-13}$,   the leading contribution to $d$ comes from the strong sector. 
On the other hand, for smaller values of $\theta$, $d$ is dominated by the weak interactions. 
In this work, we shall assume the first scenario and in the remaining part of this section we compute the QCD contribution to $d$.

Let us consider the response of the vacuum to an external electro-magnetic field $F^\textrm{ext.}_{\mu\nu}=({\bf E},{\bf B})$.
It is convenient to introduce the electric displacement field ${\bf D}$ and the auxiliary magnetic field ${\bf H}$, defined as
\be
{\bf D} &=& \frac{\partial\mathcal{L}}{\partial{\bf E}}={\bf E}+{\bf P},\\
{\bf H} &=& -\frac{\partial\mathcal{L}}{\partial{\bf B}}={\bf B}-{\bf M}.
\ee
where ${\bf P}$ and ${\bf M}$ are the polarization and magnetization vectors, respectively.
From  Eq.~(\ref{EH}) we obtain
\be
{\bf D}=[1 - 4bX - 2 dY]{\bf E} + [ 2 c Y + d X]{\bf B} + \ldots,\\ 
{\bf H}=[1 - 4bX - 2 d Y]{\bf B} - [2 c Y + d X]{\bf E} + \ldots.
\ee

Some observations on these Eq.s are in order. As expected, the non-linear vacuum polarization is a purely quantum effect. In the absence of CP violation --- i.e. if $d=0$ --- a purely electric (magnetic) external  field 
would induce a purely electric polarization (magnetization) along the same  direction,  i.e.
\be
{\bf B} =0, {\bf E} \ne 0   \stackrel{\textrm{CP}}{\Longrightarrow}  {\bf D}=  4 b E^2 {\bf E},\\
{\bf E} = 0, {\bf B} \ne 0  \stackrel{\textrm{CP}}{\Longrightarrow}  {\bf H} = - 4 b B^2 {\bf B}. 
\ee

This condition is no longer verified if CP is broken by the microscopic dynamics buried inside the quantum loops. In fact, if $d\ne 0$,  a purely magnetic (electric) external field can generate an electric polarization (magnetization) along the same direction, i.e.
\be
\label{dB}
{\bf B} \ne 0, {\bf E} = 0   \stackrel{\textrm{CP-viol.}}{\Longrightarrow} {\bf D}=  d B^2 {\bf B},\\
\label{dE}
{\bf E} \ne 0, {\bf B} = 0    \stackrel{\textrm{CP-viol.}}{\Longrightarrow} {\bf H} = d E^2 {\bf E}. 
\ee

The relationship (\ref{dB}) can be used to compute the effective coefficient of the leading CP-odd term in the effective Lagrangian (\ref{EH}). In fact, a calculation of the strong CP-odd contribution to the vacuum electric dipole moment induced by an external magnetic field was recently carried out by the authors in \cite{noi}. We have found that, for a constant and uniform external field, the vacuum electric dipole moment reads
\be
\label{VEDM}
{\bf D}=\frac{5}{9}\frac{\alpha^2 \theta}{m^2_{\eta'} (4\pi f_\pi)^2}~B^2~{\bf B},
\ee
 This result was obtained considering two quark flavors and, it corresponds to the leading-order in the chiral and $1/N_c$ expansion. Some astrophysical implications of this effect for the physics of magnetar (i.e. neutron star with extremely large magnetic fields, of the order of $10^{15}$~Gauss) have been discussed in \cite{noi}. 

The strong contribution to the coefficient $d$ is immediately obtained by matching Eq. (\ref{VEDM}) with Eq. (\ref{dB}) and reads
\be
\label{dQCD}
d=\frac{5}{9}\frac{\alpha^2 \theta}{m^2_{\eta'} (4\pi f_\pi)^2}.
\ee
We also note that using eq. (\ref{dE}) and (\ref{dQCD}), it is immediate to obtain the CP-odd magnetization induced by an external electric field
\be
{\bf M}=- \frac{5}{9}\frac{\alpha^2 \theta}{m^2_{\eta'} (4\pi f_\pi)^2}~E^2~{\bf E}
\ee

In the next section, we evaluate the contribution of the  CP-odd component of the low-energy photon-photon interaction (\ref{EH}) to the induced ellipticity of an electromagnetic wave, propagating in a resonant cavity.

\section{Vacuum Birefringence}
\label{compare}

Let us consider an electromagnetic wave,  characterized by the fields $({\bf e},{\bf b})$, propagating along the  $\hat{z}$ direction and linearly polarized along an axis rotated by
 an angle $\alpha_0$ with respect to the $\hat{y}$-axis (see Fig.~\ref{default}):
\be
{\bf e}({\bf x}, t) =  e_0~ \left ( \cos \alpha_0~ \hat{y}  + \sin \alpha_0 ~\hat{x} \right)~ e^{i  (k_z  z- \omega t)}.
\ee
The wave enters a cavity  filled with a constant and uniform magnetic field  ${\bf B}$, directed along the $\hat{y}$ axis.
 After traveling for a distance $L$,  the electromagnetic wave develops an elliptic polarization, which has both CP-even and CP-odd components. Let us first review the calculation of the CP-conserving contribution~\cite{CPevenPsi}. 

\subsection{CP-Even Vacuum Birefringence}
The dynamics of the electromagnetic field is conveniently described in terms of the vacuum electric and magnetic permeability tensors $\varepsilon_{i j}$ and $\mu_{i j}$, which are defined as 
\be
d_i=\varepsilon_{ij} \,e_j, \qquad h_i=\mu_{ij} \,b_j,
\label{permeabilities}
\ee
where  and ${\bf d}$ and ${\bf h}$ are the usual displacement vector and auxiliary magnetic field,
\be
{\bf d}=\frac{\partial\mathcal{L}}{\partial{\bf e}},
\qquad {\bf h}=-\frac{\partial\mathcal{L}}{\partial{\bf b}}.
\ee
In terms of the CP-even coefficients of the effective Lagrangian (\ref{EH}), the electric and magnetic permeability tensors have the following expression 
\be 
\label{w1}
\varepsilon_{ij}=\left(1-4bB^2\right)\delta_{ij}+2cB_iB_j,\\
\label{w2}
\mu_{ij}=\left(1-4bB^2\right)\delta_{ij}-8bB_iB_j.
\ee

Since ${\bf d}$ and ${\bf h}$ must satisfy a wave equation, $\varepsilon$ and $\mu$ must depend only on the external magnetic field ${\bf B}$ and not fields ${\bf e}$ and ${\bf b}$. Maxwell's Eq.s imply
\be
{\bf k}\cdot{\bf b} &=& 0\\
{\bf k}\cdot{\bf d} &=& 0\\
\label{eq1}
{\bf k}\times{\bf e} &=& \omega{\bf b}\\
{\bf k}\times{\bf h} &=& -\omega{\bf d},
\label{eq2}
\ee
where ${\bf k}$ is the propagation vector of the wave.  By substituting ${\bf b}$ of Eq.(\ref{eq1}) in Eq.(\ref{eq2}) we obtain 
\be
\label{eq3}
\epsilon_{i j q} ~k_j [\mu_{q l} ~({\bf k} \times {\bf e})_l] + \omega^2 \varepsilon_{i l} e_l  = 0,
\label{mat}
\ee
where  $\epsilon_{i j q}$ denotes the usual rank-3 completely anti-symmetric tensor.
Using the relationships (\ref{w1}) and (\ref{w2}),  Eq. (\ref{mat}) can be written in the following matrix form
\be 
\left(\begin{matrix}
\lambda_1 & 0 \\
0  & \lambda_2 
\end{matrix}\right)
\,\,\left(\begin{matrix}
 e_1\\
e_2
\end{matrix}\right)
=0,
\ee
with 
\be
\label{l1}
\lambda_1 &=&-\frac{{\bf k}^2}{\omega^2}[1 -12\, b~ {\bf B}^2]+1-4  b  {\bf B}^2\\ 
\label{l2}
\lambda_2 &=&-\frac{{\bf k}^2}{\omega^2}[1 - 4\,b  ~{\bf B}^2]+1+2( c - 2 b ) {\bf B}^2.
\ee
The two solutions of this equation give the refraction indeces  along $\hat{x}$ and $\hat{y}$, respectively:
\be
n_{1} &=& 1+4\, b ~{\bf B}^2 + \ldots\\
n_{2} &=& 1+c~ {\bf B}^2 + \ldots.
\label{mat1}
\ee
If the incoming wave is linearly polarized along the eigenvectors of $\lambda_1$ and $\lambda_2$ ---i.e. along the $\hat{x}$ and $\hat{y}$ axis--- it  will remain linearly polarized, even inside the region permeated by the magnetic field. On the other hand, if the incoming wave is polarized along an axis forming a finite angle $\alpha_0$ with respect to the direction selected by the eigenvector with the greatest eigenvalues ---i.e. the $\hat{y}$ axis---,  it will acquire an ellipticity \cite{CPevenPsi}
\be\label{esteven}
\Psi_\textrm{CP-even}({\bf B}, L)= \frac{\pi}{\lambda}   (n_2({\bf B})- n_1({\bf B})) L \times \sin 2 \alpha_0 = \frac{\pi}{\lambda} ~ (c-4 b ) ~{\bf B}^2~ L \sin 2 \alpha_0.
\ee

\subsection{CP-odd Vacuum Birefringence}\label{calccpodd}

We now compute the ellipticity induced on the electromagnetic wave, in the presence of CP-violating vacuum polarization. 
In this case, the electric displacement field ${\bf d}$ and auxiliary magnetic field ${\bf h}$ 
receive additional contributions $\Delta{\bf d}$ and $\Delta {\bf h}$, which depend on the CP-odd effective coefficient $d$,
\be
{\bf d} &=& {\bf d}_\textrm{CP-even}+ \Delta{\bf d}\\
{\bf h} &=& {\bf h}_\textrm{CP-even}+ \Delta{\bf h},
\ee 
with
\be
\Delta {\bf d} &=& 2d\,{\bf B}[{\bf b} \cdot {\bf B}]+d\,{\bf b}\,{\bf B}^2\\
\Delta {\bf h} &=& -2d\,{\bf B}[{\bf e}\cdot{\bf B}]-d\,{\bf e}\,{\bf B}^2.
\ee
Maxwell's Eq.s now give
\be\label{eq3CP}
\epsilon_{i j q} ~k_j [\mu_{q l} ~({\bf k} \times {\bf e})_l] + \omega^2 \varepsilon_{i l} e_l =2d\,[(\epsilon_{ijq}k_jB_q) e_l B_l+B_i([({\bf k}\times{\bf B})_je_j].
\ee 
The CP-odd interactions introduce non-diagonal elements in the matrix (\ref{mat}), which now becomes:
\be
\label{mat2} 
\left(\begin{matrix}
\lambda_1 & C \\
C  & \lambda_2 
\end{matrix}\right)
\,\,\left(\begin{matrix}
 e_1\\
e_2
\end{matrix}\right)
=0,
\ee
where $\lambda_1$ and $\lambda_2$ are the same as in Eq.s (\ref{l1}),(\ref{l2}) and
\be
C= 2 \, \frac k \omega d\, {\bf B}^2.
\ee
As in the CP-even case, an incoming wave which is  linearly polarized along the eigenvectors of the matrix (\ref{mat2}) does not acquire ellipticity. Up to higher corrections in the effective coefficient $d$, the eigenvalues of (\ref{mat2}) --- i.e. the refraction indexes--- are the same as those of (\ref{mat1}).  On the other hand, the new eigenvectors are rotated by an angle $\beta$ given by
\be
\beta=\frac{d}{c-4b}+O({\bf B}^2).
\ee
Consequently, the final expression for the ellipticity acquired in the cavity, in the presence of CP-violation reads
\be
\Psi =  \frac{\pi}{\lambda} ~ (c-4 b ) {\bf B}^2~ L \sin 2 (\alpha_0-\beta).
\ee
We expect CP-violation to provide at most a small correction to the total ellipticity, i.e. $\beta \ll \alpha_0$. In this case,  
\be
\Psi \simeq \Psi_\textrm{CP-even} +  \Psi_\textrm{CP-odd},
\ee
with
\be\label{estodd}
\Psi_\textrm{CP-odd}=-2 \pi d \frac L \lambda {\bf B}^2 \cos 2 \alpha_0.
\label{CPodd}
\ee
This Eq. has been independently obtained in \cite{Liao1} and \cite{Rizzo}.

Within the Standard Model, the CP-odd correction is   indeed extremely small. For example, if we  consider a set-up characterized by  an angle $\alpha_0=\pi/8$, for which the trigonometric factor cancels out, then the ratio between the  $\Psi_\textrm{CP-even}$ and $\Psi_\textrm{CP-odd}$ is
\be
\left(\frac{\Psi_\textrm{CP-odd}}{|\Psi_\textrm{CP-even}|}\right)_{\alpha=\pi/8} \simeq \frac{2d}{c-4 b} \simeq \theta \times 10^{-12}< 10^{-22}.
\ee


However, it is important to stress that the CP-odd and CP-even contributions to the ellipticity are {\it qualitatively} different and therefore can be experimentally disentangled. In particular, if $\alpha_0$ is chosen to be $0,\pi/2,\ldots$, then the CP-even contribution vanishes and the entire ellipticity is due to CP-violating quantum photon-photon interactions. In this case, any significant deviation from our estimate, 
\be
\Psi_\textrm{CP-odd}\simeq -\frac{10 \pi \alpha^2 \theta}{9 m^2_{\eta'} (4\pi f_\pi)^2}  \frac L \lambda {\bf B}^2,
\label{PsiODD}
\ee
would represent a clean signature of CP-violating physics beyond the Standard Model. 

In the next section, we shall discuss the possibility of measuring the CP-odd ellipticity using a Fabry-Perot cavity.

\section{Vacuum Birefringence in a Resonant Cavity}\label{nuovo}

We now imagine that the electromagnetic wave propagating in the cavity is reflected by a mirror. After the reflection, both the polarization vector {\bf e} and the wave vector {\bf k} change sign: this affects both the angles $\alpha_0$ and $\beta$. In fact, the angle $\alpha_0$ will undergo a rotation of $180^\circ$
\be
\alpha_0&&\longrightarrow\alpha^{'}_0=\alpha_0+\pi\\
\beta&&\longrightarrow\beta^{'}=-\beta,
\ee
while the angle $\beta$ changes sign. The latter effect occurs because $\beta$ is proportional to the coefficient $C$ in Eq.(\ref{mat2}): since $C$ is linear in $k$, by inverting the direction of the electromagnetic wave the sign of $\beta$ changes. On the contrary, the indexes of refraction $n_1$ and $n_2$ remain unchanged, because $\lambda_1$ and $\lambda_2$ are quadratic in $k$, see Eq.s(\ref{l1}) and (\ref{l2}).\\
Thus, the contribution to the total ellipticity during the return part of the round trip is given by
\be
\Psi^{'} &=&  \frac{\pi}{\lambda} ~ (c-4 b ) {\bf B}^2~ L \sin 2 (\alpha^{'}_0-\beta^{'})\\
 &=&  \frac{\pi}{\lambda} ~ (c-4 b ) {\bf B}^2~ L \sin 2 (\alpha_0+\beta)
\label{CPodd}
\ee
Hence, the total amount of CP-odd ellipticity of the wave after a round trip is
\be
\Psi_\textrm{CP-odd,R}=\Psi_\textrm{CP-odd}+\Psi^{'}_\textrm{CP-odd}=0
\ee
After being reflected 2N times in the resonant cavity, the total ellipticity gained by the electromagnetic wave will be
\be
\Psi_{TOT}=(2N+1)\Psi_\textrm{CP-even}+\Psi_\textrm{CP-odd}
\ee
Thus, we conclude that the experimental setup formed by a magnetic field in a resonant cavity is not suitable for measurements of CP-odd ellipticity: there is no enhancement of the CP-odd signal after several round trips.\\

We now show that if the cavity is also permeated by an electric field set up in the plane perpendicular to ${\bf k}$, the CP-odd ellipticity does not cancel out, in general, after a round trip. To this end, we performed again the calculation in section \ref{calccpodd} considering an external electromagnetic field $({\bf E},{\bf B})$, instead of only $({\bf 0},{\bf B})$. After a tedious but straightforward calculation one obtains the following results for the {\it forward} and 
{\it backward} ellipticities $\Psi^f$ and $\Psi^b$:
\be
\Psi^{f}_\textrm{CP-even}&=&\frac{\pi}{\lambda}(c-4b)\left[{\bf B}^2\times\sin2\alpha_B-{\bf E}^2\times\sin2\alpha_E-2|{\bf E}|\,|{\bf B}|\cos(\alpha_B+\alpha_E)\right]\\
\Psi^{f}_\textrm{CP-odd}&=&-2d\frac{\pi}{\lambda}\left[{\bf B}^2\times\cos2\alpha_B-{\bf E}^2\times\cos2\alpha_E+2|{\bf E}|\,|{\bf B}|\sin(\alpha_B+\alpha_E)\right]\\
\Psi^{b}_\textrm{CP-even}&=&\frac{\pi}{\lambda}(c-4b)\left[{\bf B}^2\times\sin2\alpha_B-{\bf E}^2\times\sin2\alpha_E+2|{\bf E}|\,|{\bf B}|\cos(\alpha_B+\alpha_E)\right]\\
\Psi^{b}_\textrm{CP-odd}&=&2d\frac{\pi}{\lambda}\left[{\bf B}^2\times\cos2\alpha_B-{\bf E}^2\times\cos2\alpha_E-2|{\bf E}|\,|{\bf B}|\sin(\alpha_B+\alpha_E)\right].
\ee
After a round trip, the total ellipticity amounts to
\be
\Psi_\textrm{CP-even,R}&=&\Psi^{f}_\textrm{CP-even}+\Psi^{b}_\textrm{CP-even}=2\frac{\pi}{\lambda}(c-4b)\left[{\bf B}^2\times\sin2\alpha_B-{\bf E}^2\times\sin2\alpha_E)\right]\label{cpevenref}\\
\Psi_\textrm{CP-odd,R}&=&\Psi^{f}_\textrm{CP-odd}+\Psi^{b}_\textrm{CP-odd}=-8d\frac{\pi}{\lambda}|{\bf E}|\,|{\bf B}|\sin(\alpha_B+\alpha_E).\label{cpoddref}
\ee
Hence after 2N reflections, the total ellipticity is
\be
\Psi_\textrm{TOT}=N\left[\Psi_\textrm{CP-even,R}+\Psi_\textrm{CP-odd,R}\right]+\left[\Psi^f_\textrm{CP-even}+\Psi^f_\textrm{CP-odd}\right]
\ee
Notice that the CP-odd ellipticity is maximum if $(\alpha_B+\alpha_E)=~\pi/2,3\pi/2,..$ and vanishes for $(\alpha_B+\alpha_E)=0,\pi,..$.
Notice also that, if the electric and magnetic fields are orthogonal to each other --- i.e. if $\alpha_E-\alpha_B=\pm\pi/2$---, then one recovers the structure of  the CP-even and CP-odd ellipticities  in Eq.s(\ref{esteven}) and (\ref{estodd})
\be
\Psi_\textrm{CP-even}&=&2\frac{\pi}{\lambda}(c-4b)\left[{\bf B}^2+{\bf E}^2\right]\sin2\alpha_B\\
\Psi_\textrm{CP-odd}&=&\mp8d\frac{\pi}{\lambda}|{\bf E}|\,|{\bf B}|\cos2\alpha_B
\ee
On the contrary, if the electric and magnetic fields are parallel to each other --- i.e. $\alpha_B=\alpha_E$--- both  contributions will be proportional to $\sin2\alpha_B$
\be
\Psi_\textrm{CP-even}&=&2\frac{\pi}{\lambda}(c-4b)\left[{\bf B}^2-{\bf E}^2\right]\sin2\alpha_B\\
\Psi_\textrm{CP-odd}&=&-8d\frac{\pi}{\lambda}|{\bf E}|\,|{\bf B}|\sin2\alpha_B.
\ee
In the next section we study the effects of modulating the external fields on the spectrum of the intensity of the outgoing wave.

\section{Implications on the PVLAS-Phase II, and BMV Experiments}
\label{pvlassect}

The main goal of the experiments PVLAS Phase-II \cite{PVLASII} and BMV \cite{Rizzo2} is to study the birefringence and the dichroism induced by the quantum fluctuations in the QED vacuum, in the presence of an external magnetic field {\bf B}. In the following, we restrict our attention to the implications of CP-violating quantum fluctuations on the vacuum birefringence. We compare two scenarios, one in which only the magnetic field permeates the cavity and one in which also an electric field {\bf E} is present.\\

\begin{figure}[t!]
\begin{center}

\includegraphics[width=14 cm]{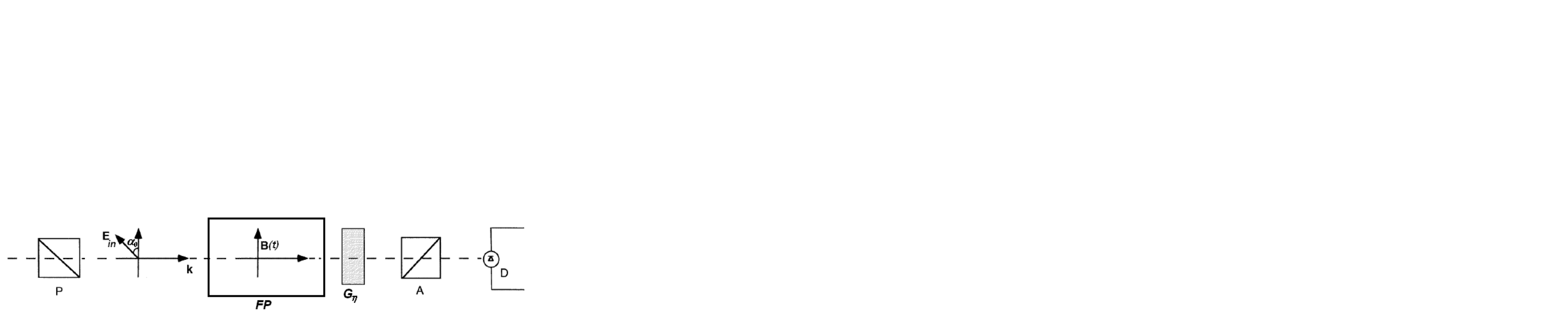}
\caption{Schematic representation of the experimental setup of PVLAS (see text). $P$ and $A$ are the cross-polarizers, $FP$ is the Fabry-Perot cavity, $G_\eta$ is generator of the additional ellipticity $\eta(t)$ and $D$ is the detector.}
\label{polarization}
\end{center}
\end{figure}

Let us begin with the first case, which is the one implemented at PVLAS and BMV. The experimental setup consists of a Fabry-Perot cavity, placed between two crossed polarizers (~Fig.~\ref{polarization}~ ). The cavity is uniformly filled with a magnetic field ${\bf B}$, rotating with angular frequency $\omega_B$, along the plane perpendicular to the wave vector of the incoming light.
Since the  magnetic field is rotating, the total induced ellipticity is time-dependent, $\Psi=\Psi(\omega_B t + \alpha_B)$, 
where $\alpha_B$ is the angle between the polarization vector and the magnetic field at the time $t=0$. 

It can be shown  that the intensity of the outgoing wave has the following form~\cite{PVLASII, outgoing}
\begin{equation}\label{Iout}
I_{out}~=~I_{in}~\left|\ \frac{2\mathcal{F}}{\pi}~ i~\Psi_\textrm{CP-even}\left(\omega_{B}\,t+\alpha_B\right)+i~\Psi_\textrm{CP-odd}\left(\omega_{B}\,t+\alpha_B\right)\right|^2,
\end{equation}
where $I_0$ is the intensity of the incoming wave, $\mathcal{F}$ is the so-called finesse of the Fabry-Perot cavity.

The ellipticity $\Psi$ generated by the quantum vacuum polarization is in general expected to be a small effect. 
Hence, in order to increase the intensity of the wave coming out the last polarizer, an additional time-dependent classical ellipticity $\eta(t)$ is introduced by means of a modulator. This way, the intensity of the outgoing wave becomes linear in the ellipticity $\Psi(t)$:
\begin{eqnarray}\label{powerspectrum}
I_{out}
\simeq I_{in} \eta(t)~\left\{\eta(t)+4\frac{\mathcal{F}}{\pi}~\Psi_\textrm{CP-even}\left(\omega_{B}\,t+\alpha_0\right)+2~\Psi_\textrm{CP-odd}\left(\omega_{B}\,t+\alpha_B\right)\right\}+O\left(\psi^2\right).
\end{eqnarray}

In particular, let us consider the case in which the classical ellipticity $\eta(t)$ is modulated with a frequency $\omega_{\eta}$, 
\be
\eta(t)=\eta_0\,\cos(\omega_{\eta}\,t+\alpha_{\eta}).
\ee 
In this case, the intensity of the outgoing wave is
\begin{eqnarray}\label{intout}
I_{out,\psi}(t)&=&  L\frac{\pi}{\lambda}\,{\bf B}^2\, I_{in}\,\eta_0 \Big[2\frac{\mathcal{F}}{\pi}
(c-4b) \left\{\sin\left[(\omega_{\eta}+2\omega_{B})t+\alpha_{\eta}+2\alpha_0\right]-\sin\left[(\omega_{\eta}-2\omega_{B})t+\alpha_{\eta}-2\alpha_0\right]\right\}+\nonumber\\
&-& 2 d\left\{\cos\left[(\omega_{\eta}+2\omega_{B})t+\alpha_{\eta}+2\alpha_0\right]+\cos\left[(\omega_{\eta}-2\omega_{B})t+\alpha_{\eta}-2\alpha_0\right]\right\}\Big].
\end{eqnarray}

The CP-even and CP-odd contributions to the induced ellipticity can be disentangled by Fourier analysis.  In fact, we see that the Fourier spectrum of $I_{out,\psi}$ contains:
\begin{itemize}
\item Two CP-even Fourier components with  amplitude given by
\be\label{aevenB}
A_{even}=2(c-4b)\,\frac{L}{\lambda}\,{\bf B}^2\, I_0\,\eta_0~\mathcal{F}
\ee
and frequency and phase given by
\be
\Omega_{even,\pm}=\omega_{\eta} \pm  2\omega_B; ~\quad \Phi_{even,\pm}=\alpha_{\eta} \pm 2\alpha_0
\ee
\item Two CP-odd Fourier components of amplitude given by
\be\label{aoddB}
A_{odd}=2\,d\,L\frac{\pi}{\lambda}\,{\bf B}^2\,I_0\,\eta_0
\ee
and frequency and phase given by
\be
\label{peaks}
\Omega_{odd,\pm}=\omega_{\eta} \pm 2 \omega_B;~\quad\Phi_{odd,\pm}=\alpha_{\eta}\pm\left(2\alpha_0-\frac{\pi}{2}\right).
\ee
\end{itemize}
Note that, except for $\alpha_0=\frac{\pi}{8}$, the phases of the CP-even and CP-odd components are  different. This fact implies that, at least in principle, it should be possible to probe 
directly the CP-odd part of the photon-photon interaction by analyzing the spectrum of the outgoing wave. Unfortunately, as we noticed in section \ref{nuovo}, the amplitude (\ref{aoddB}) of the CP-odd ellipticity is suppressed by a factor $\mathcal{F}\propto N$, where N is the number of round trips in the cavity, respect to the CP-even amplitude (\ref{aevenB}): any signal of CP-violation would be hardly measurable.

We now discuss what would happen if a time dependent electric field is placed in the Fabry-Perot cavity, along with the magnetic field.
If the electric field is modulated with frequency $\omega_E$ and phase $\alpha_E$, by replacing Eq.s(\ref{cpevenref}) and (\ref{cpoddref}) in Eq.(\ref{Iout}) the intensity of the outgoing wave in Eq.(\ref{intout}) gives, up to $O(\mathcal{F}^0)$
\be\label{intout}
I_{out,\psi}(t)&=& 2\frac{L }{\lambda}\, I_{in}\,\eta_0 \,\mathcal{F}\Big[(c-4b){\bf B}^2 \left\{\sin\left[(\omega_{\eta}+2\omega_{B})t+\alpha_{\eta}+2\alpha_B\right]-\sin\left[(\omega_{\eta}-2\omega_{B})t+\alpha_{\eta}-2\alpha_B\right]\right\}+\nonumber\\
&-&(c-4b){\bf E}^2 \left\{\sin\left[(\omega_{\eta}+2\omega_{E})t+\alpha_{\eta}+2\alpha_E\right]-\sin\left[(\omega_{\eta}-2\omega_{E})t+\alpha_{\eta}-2\alpha_E\right]\right\}+\nonumber\\
&-& 4 d|{\bf E}|\,|{\bf B}|\left\{\sin\left[(\omega_{\eta}+\omega_{B}+\omega_E)t+\alpha_{\eta}+\alpha_B+\alpha_E\right]-\sin\left[(\omega_{\eta}-\omega_{B}-\omega_E)t+\alpha_{\eta}-\alpha_B-\alpha_E\right]\right\}\Big].
\ee
Thus, the spectrum of the outgoing intensity has six characteristic Fourier components
\begin{itemize}
\item Two CP-even Fourier components with amplitude given by
\be\label{aeven}
A_{even,B}=2(c-4b)\,\frac{L}{\lambda}\,{\bf B}^2\, I_0\,\eta_0~\mathcal{F}
\ee
and frequency and phase given by
\be
\Omega_{even,\pm}=\omega_{\eta} \pm  2\omega_B; ~\quad \Phi_{even,B,\pm}=\alpha_{\eta} \pm 2\alpha_B
\ee
\item Two CP-even Fourier components with amplitude given by
\be\label{aeven}
A_{even,E}=2(c-4b)\,\frac{L}{\lambda}\,{\bf E}^2\, I_0\,\eta_0~\mathcal{F}
\ee
and frequency and phase given by
\be
\Omega_{even,\pm}=\omega_{\eta} \pm  2\omega_E; ~\quad \Phi_{even,E,\pm}=\pi+\alpha_{\eta} \pm 2\alpha_E
\ee
\item Two CP-odd Fourier components of amplitude given by
\be\label{aodd}
A_{odd}=8\,d\,\frac{L}{\lambda}\,|{\bf E}|\,|{\bf B}|\,I_0\,\eta_0\mathcal{F}
\ee
and frequency and phase given by
\be
\label{peaks}
\Omega_{odd,\pm}=\omega_{\eta} \pm ( \omega_B+\omega_E);~\quad\Phi_{odd,\pm}=\pi+\alpha_{\eta}\pm\left(\alpha_B+\alpha_E)\right).
\ee
\end{itemize}
Based on the considerations made in section \ref{nuovo}, we realize that for an experimental setup in which $\alpha_{E}-\alpha_B=\pi/2$ and $\omega_E=\omega_B$, the 
four CP-even peaks merge into  two peaks. The  CP-odd Fourier components of the outgoing intensity spectrum have the same frequency of the CP-even ones, 
but their phases are shifted by $\pi/2$. Hence, we have recovered our previous result with
\be
(c-4b){\bf B}^2&&\longrightarrow(c-4b)\left[{\bf B}^2+{\bf E}^2\right]\\
d{\bf B}^2&&\longrightarrow\frac{4}{\pi}d\,|{\bf E}|\,|{\bf B}|\,\mathcal{F}
\ee

The presence of an electric field in the cavity can significantly increase the sensitivity of the experimental set-up to the CP-odd ellipticity only 
 if the round-trip contribution $\propto d |{\bf E}|\,|{\bf B}| \mathcal{F}$ 
is much larger than the  one-way contribution, $\propto d{\bf B}^2$. 
This condition is verified if 
\be\label{cond}
4{\bf E}|\,\mathcal{F}\gg\pi{\bf B}.
\ee
For example, for a typical finesse of $10^5$,  the condition (\ref{cond}) holds for electric fields  $\gg 10V/cm$.

The contribution to the coefficient $d$ coming form the $\theta$-term of QCD is at least 20 order of magnitudes smaller than the coefficient of the CP-even term $c-4b$.
 Given such a huge difference, it is difficult to imagine that an experiment with a resonant cavity will be able to reach the sensitivity required to resolve the Standard Model contribution. 
Hence,  observing evidence for the peaks (\ref{peaks}) in the spectrum  of the outgoing wave would represent  a clean signature of CP-violation coming from microscopic dynamics beyond the Standard Model.

\section{Conclusions}
\label{concl}

In this paper, we have considered CP-violation in photon-photon  interactions, at energies much below the electron mass scale. 
In such a kinematic regime, the photon dynamics can be described by an effective field theory, in which quantum loops are replaced by contact vertexes. 
We have seen that a  CP-violation in this channel would result in a specific contribution to the ellipticity acquired by an electromagnetic wave propagating in a Fabry-Perot cavity, in the presence of an external electric field along with a magnetic field. 
Interestingly, the CP-even and CP-odd ellipticities are  qualitatively distinct  and give raise to additional  Fourier components, in the power spectrum of the out-going wave.
 
We have estimated the magnitude of this effect, in a scenario in which the only CP-violation comes from the $\theta$ angle of QCD. 
To this end, we have computed the coefficient of the lowest-dimensional CP-odd operator in the photon effective field theory, at the leading order in chiral perturbation theory and in the $1/N_c$ expansion.
Given the smallness of this effect, it is difficult to imagine that PVLAS or BMV experiments will reach the sensitivity required to improve on the present upper bound for $\theta$.  On the other hand, 
we have shown that, if a large electric field is dialed into the cavity, the analysis of the spectrum of intensity of the outgoing wave may be used to set upper bounds on exotic CP-violating dynamics.

\end{document}